\documentclass[aps,twocolumn,nofootinbib,longbibliography,prd]{revtex4-1}
\usepackage[babel]{csquotes}
\usepackage{graphicx}
\usepackage{amsmath,amssymb}
\usepackage[colorlinks,citecolor=blue,linkcolor=blue,urlcolor=blue]{hyperref}
\usepackage{mathrsfs}
\usepackage{enumerate}
\def\be{\begin{equation}}
\def\ee{\end{equation}}

\begin{document}

\title{Black holes have more states \\ than those giving the Bekenstein-Hawking entropy: \\ a simple argument.}

\author{Carlo Rovelli}

\affiliation{CPT, Aix-Marseille Universit\'e, Universit\'e de Toulon, CNRS, Case 907, F-13288 Marseille, France.}
\date{\small\today}

\begin{abstract} 
\noindent  It is often assumed that the maximum number of independent states a black hole may contain is $N_{BH}=e^{S_{BH}}$, where $S_{BH}=A/4$ is the Bekenstein-Hawking entropy and $A$ the horizon area in Planck units.  I present a simple and straightforward  argument showing that the number of states that can be distinguished by local observers inside the hole must be greater than this number. 
\end{abstract} 

\maketitle

There are convincing arguments supporting the idea that the thermodynamical interactions between a black hole and its surroundings are well described by treating the black hole as a system  {with} 
\be 
N=e^{A/4}
\label{Ne}
\ee  (orthogonal) states, where $A$ is the area of the surface enclosing the black hole in Planck units $\hbar=G=c=1$.   These arguments are convincing.   However, it has since become fashionable to deduce that the black hole itself cannot have more than $e^{A/4}$ states, as the number of states on a region enclosed in a sphere of radius $A$ cannot be larger than $e^{A/4}$ (see, for instance, the discussion in \cite{don} and the references therein). Here, I present an argument indicating that this further step is wrong: the actual number $N$ of independent states enclosed inside a surface of area $A$ can be larger than $e^{A/4}$. 

The possibility of a distinction between $N$ and $e^{A/4}$ is because, according to classical general relativity, the interaction between a black hole and its surroundings is entirely determined by what happens in the vicinity of the horizon. This may be true in general; therefore, it is possible that $e^{A/4}$ counts only states that can
 be distinguishable from the exterior, which may be called ``surface'' states. These and only these may be those governing the thermodynamical interactions with the exterior. Instead, $N$ also counts states that can be distinguished by local observables \emph{inside} the horizon. Here, I argue that to have more states than $e^{A/4}$ is not just a possibility: it follows from elementary considerations \mbox{of causality}. 

\begin{figure}
\center{\includegraphics[height=8cm]{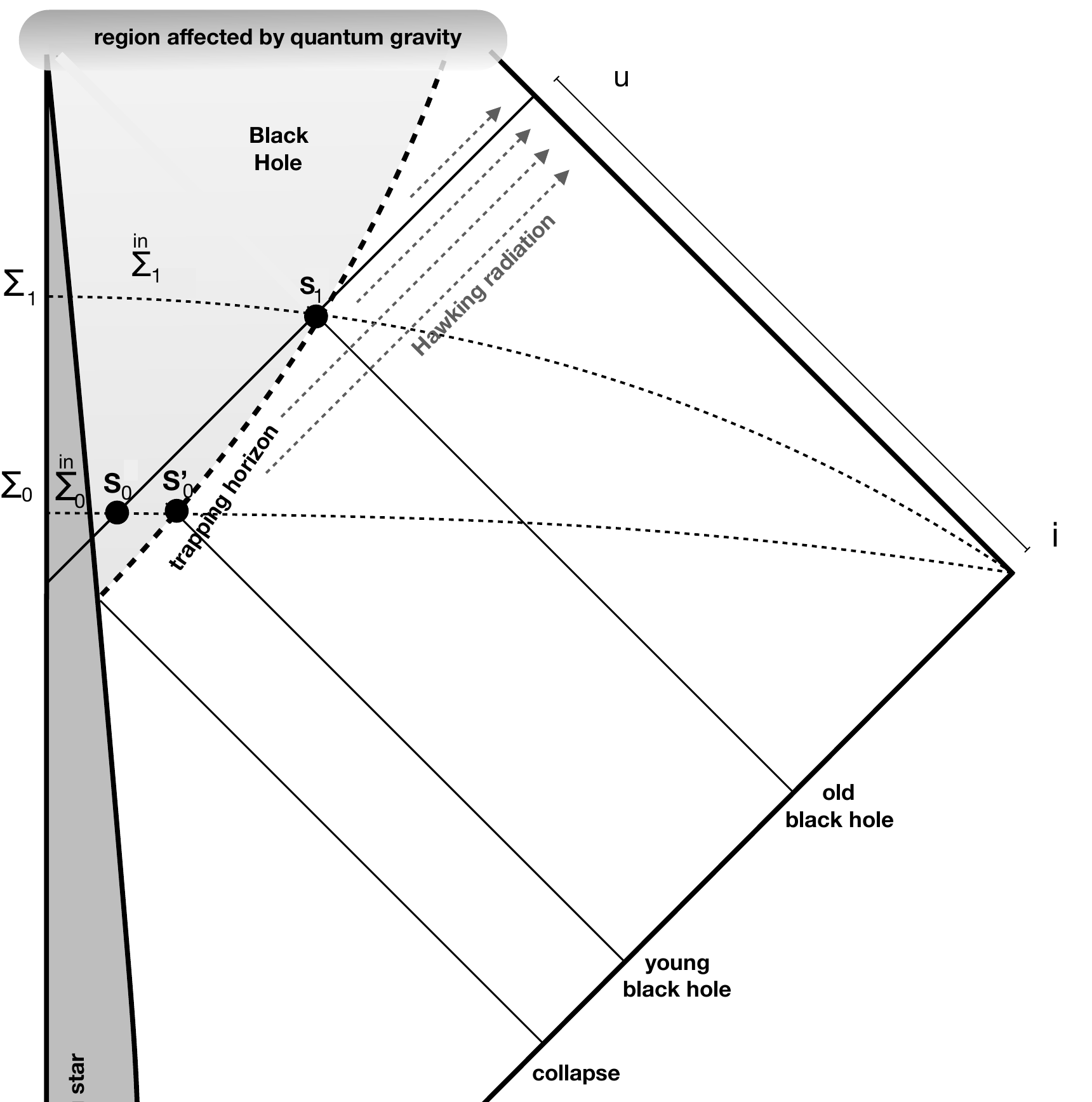}}\vspace{6mm}
\vspace{-10pt}\caption{  {Conformal} 
 diagram of a gravitational collapse followed by Hawking evaporation. The dark grey region is the collapsing object, the light grey region is the black hole. The dotted line is the trapping horizon. The two Cauchy surfaces used in the paper are the light dashed lines.}
\label{figuno}
\end{figure}

To show this, consider a gravitationally collapsed object. See Figure \ref{figuno}.  Let $\Sigma_0$ be a Cauchy surface that crosses the horizon but does not hit the singularity.  Let $\Sigma_1$ be a later similar Cauchy surface and $i=0,1$.  We disregard what happens in the region with a high curvature; then, the event horizon may not be defined \cite{BHbounce}, as the curvature becomes Planckian outside the hole before the end of the evaporation. However, a quasi-local horizon such a trapping horizon is defined.  In the presence of the back reaction of the Hawking radiation, this horizon can be timelike.    Let $S_0'$ be the intersection of $\Sigma_0$ with the trapping horizon and $S_1$ be the intersection of $\Sigma_1$ with the trapping horizon. Furthermore, let $\Sigma_1^{in}$ be the open portion of $\Sigma_1$ inside the $S_1$, let $\Sigma_0^{in}$ be the intersection of $\Sigma_0$ with the past causal domain of dependence of $\Sigma_1^{in}$, and let $S_0$ its boundary.  See the figure for an easy identification of these geometrical locuses. Finally, let $A_i$ and $A_0'$ be the areas of the spheres $S_i$ and $S_0'$, respectively. Assume that no positive energy falls into the horizon during the interval between the two surfaces.  Let quantum fields live on this geometry, back-reacting on it \cite{qft}.

Because of the back-reaction of the Hawking radiation, the area of the horizon shrinks; therefore  
\be
             A_1 <  A'_0. 
            \label{unoprimo}
\ee
If the hole is macroscopic, the difference between $A'_0$ and $A_0$ is of the order of a Planck area. To the best of my knowledge, this was first shown by York in \cite{1} (see Equations (2.14) and (2.15) and the surrounding text), where the region between the null and timelike surface  is called the `quantum ergosphere'. (I thank Eugenio Bianchi for tracing this reference.)  It follows that if the black hole has evaporated a macroscopic portion of its mass, we also have 
\be
             A_1 <  A_0. 
            \label{uno}
\ee
Consider the evolution of the quantum fields from $\Sigma_0$ to $\Sigma_1$. This is a region  away from the singularity and from the high curvature portion of spacetime; we can therefore assume conventional quantum field theory to hold here, without  strange quantum gravity effects, at least up to high energy scales.  By construction, $\Sigma_0^{in}$ is in the causal past of $\Sigma_1^{in}$.  This implies that any local observable on $\Sigma_0^{in}$ is fully determined by observables on  $\Sigma_1^{in}$.  That is, if ${\cal A}_i$ is the local algebra of observables on  $\Sigma_i^{in}$, then ${\cal A}_0$ is a subalgebra of ${\cal A}_1$:
\be
{\cal A}_0\subset {\cal A}_1.
\ee
 {Therefore}, any state on ${\cal A}_1 $ is also a state on ${\cal A}_0$, and if two such states can be distinguished by observables in ${\cal A}_0$, they certainly can be distinguished by observables in ${\cal A}_1$ as the first are included in the latest. Therefore, the states that can be distinguished by ${\cal A}_0$ ---which is to say: on $\Sigma_0^{in}$--- can also be distinguished by ${\cal A}_1$ ---which is to say: on $\Sigma_1^{in}$.  Therefore, the distinguishable (orthogonal) states on $\Sigma_0^{in}$ are a subset of those in $\Sigma_1^{in}$.  

How many are there? Either they are an infinite number, or a finite number due to some high-energy (say Planckian) cut-off. If there is an infinite number of (orthogonal) states, then immediately the number of states distinguishable from inside the black hole is larger that $N_{BH}$, which is finite. So, in this case, we immediately have what we wanted. If instead there is a finite number, then the number $N_1$ of distinguishable states on $\Sigma_1^{in}$ must be equal to or larger than the number $N_0$ of states distinguishable on $\Sigma_0^{in}$, because the second is a subset of the first. That is
\be
            N_1\ge N_0. 
            \label{due}
\ee
 {Comparing} Equations \eqref{Ne}, \eqref{uno}, and \eqref{due} immediately shows that it is impossible that $N_i=e^{A_i/4}$, as the exponential is a monotonic function.  

The conclusion is that the number of states distinguishable from the interior of the black hole must be larger than the number $e^{A/4}$ of states contributing to the Bekenstein--Hawking entropy. As the second shrinks to zero with evaporation, the first must overcome the second at some point. Therefore, in the interior of a black hole, there are more possible states than $e^{A/4}$.  

The physical interpretation of the conclusion is simple: the thermal behaviour of the black hole described by the Bekenstein--Hawking entropy $S=A/4$ is  determined by the physics of the vicinity of the horizon, not by the states in the interior. 

In classical general relativity, the effect of a black hole on its surroundings is independent from the black hole interior.  A vivid expression of this fact is in the numerical simulations of black hole merging and radiation emission by oscillating black holes: in writing the numerical code, it is routine to  cut away a region inside the (trapping) horizon: it is irrelevant for whatever happens outside. This is true in classical general relativity, and there is no compelling reason to suppose it to fail if quantum fields are around. Therefore, a natural interpretation of $S_{BH}={A/4}$~ is to count states of near-surface degrees of freedom, not interior ones.  

The idea that \emph{only} surface states are relevant for the Bekenstein-Hawking entropy is of course not new: it has a long  history  \cite{1,2,3,4,5,6,7,8,9}. See, in \mbox{particular, \cite{LQG-BH,Strominger}} in support of this idea from two different research camps, loops and strings. 
 A classic discussion on the question whether the states relevant for the black hole \emph{thermodynamical} entropy are surface states or interior states is the trialogue \cite{10}.  The argument presented here strongly supports the surface states option, by making clear that there are interior states that do not affect the Bekenstein--Hawking entropy.  

This conclusion does not contradict the various arguments identifying the Bekenstein--Hawking entropy as a measure of state counting. Indeed, it is supported by the membrane paradigm \cite{membrane} and Loop Quantum Gravity \cite{LQG-BH,LQG-BH2}, which both show explicitly that the relevant states that are counted are surface states. But also the string theory counting \cite{StringBH,StringBH2} supports this conclusion, because this derivation is in a context where the relevant state space is identified with the scattering state space in the presence of an event horizon, and this state space could well be blind to interior observables.  

The consequences of this observation are far reaching for the discussions on the black hole information paradox \cite{don,thooft}. The solid version of the paradox is Page's \cite{Page}, which does not require hypotheses on the future of the hole. If there are more states available in a black hole than $e^{A/4}$, then Page's argument for the information loss paradox fails. Page's argument is indeed based on the fact that if the number of black hole states is determined by the area, then there are no more available states to be entangled with the Hawking radiation when the black hole shrinks. For the radiation to be thermal, it must be entangled with something, and the only option is earlier Hawking quanta, and this is in tension with quantum field theory.  But, if there can be many states inside a black hole with small horizon area, then late-time Hawking radiation does not need to be correlated with early time Hawking radiation, because it can simply be correlated with internal black hole states, even when the surface area of the back hole has become small. 

Recall that the interior of an old black hole can have large volume even if its horizon has a small area. It was shown in \cite{BHvolume} that at a time $v$ after the collapse, a black hole with mass $m$ has interior volume 
\be
V\sim 3\sqrt{3} \pi\; m^2 v
\ee
for $v\ll m$. See also \cite{BHvolume2, ing,yen,yen2,shao} and \cite{review} for a review.  This volume may store a large number of states.  This information can leak out, possibly slowly, from a long living remnant after the end of the evaporation \cite{BHbounce,BHbounce2}, if much of it is in long wavelength modes (see \cite{abhay, met} and the references therein). Therefore, information \emph{ {can}} emerge from the hole, before total dissipation, and is not lost. 

These observations go against diffused prejudices regarding holography, but we should not be blocked by prejudices. The result presented here does not invalidate holographic ideas: it sharpens them by pointing out that what is bound by the area of the boundary of a region is not the number of possible states in the region. It is only the number of states distinguishable from observations outside the region.

\centerline{---}

I thank Don Marolf, Tommaso De Lorenzo, Alejandro Perez and Eugenio Bianchi for crucial conversations that have lead to this result.

\end{document}